\documentclass[10pt,draft]{dis03}
\usepackage{epsf,amsmath}

\renewcommand{\baselinestretch}{1.15}
\begin{document}

\title{RUN II DIFFRACTIVE MEASUREMENTS AT CDF}
\author{KOJI TERASHI \\
(For the CDF Collaboration)\vspace{0.3cm}\\
The Rockefeller University\\
1230 York Avenue, New York, NY 10021, USA\\
E-mail: terashi@rockefeller.edu}

\maketitle

\vspace{0.2cm}
\begin{abstract}
\noindent We present results on hard diffraction obtained
by the CDF Collaboration in Run II proton-antiproton collisions at the 
Fermilab Tevatron. Run I CDF results on hard diffraction are also
reviewed.
\end{abstract}

\section{Introduction} 
\begin{figure}
\begin{center}
\centerline{\epsfxsize=12cm\epsfbox{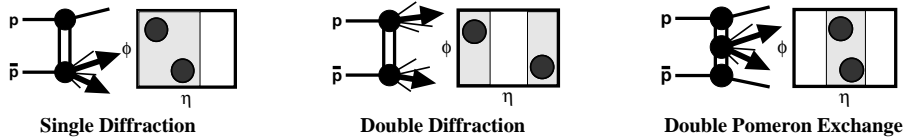}}
\renewcommand{\baselinestretch}{0.95}
\caption{\label{fig:diff_diagram}
Diagrams and event topologies of dijet production in 
single diffraction (left), double diffraction (middle) and
double pomeron exchange (right).}
\end{center}
\end{figure}

\begin{figure}
\begin{center}
\epsfxsize=3.5cm\epsfbox{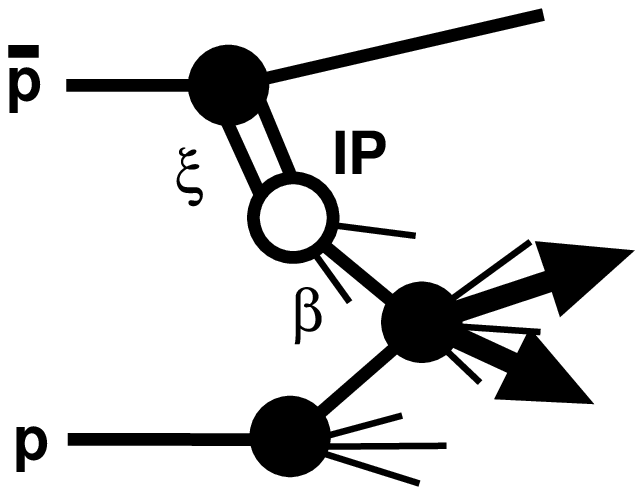}
\raisebox{-1cm}{\epsfxsize=4.3cm\epsfbox{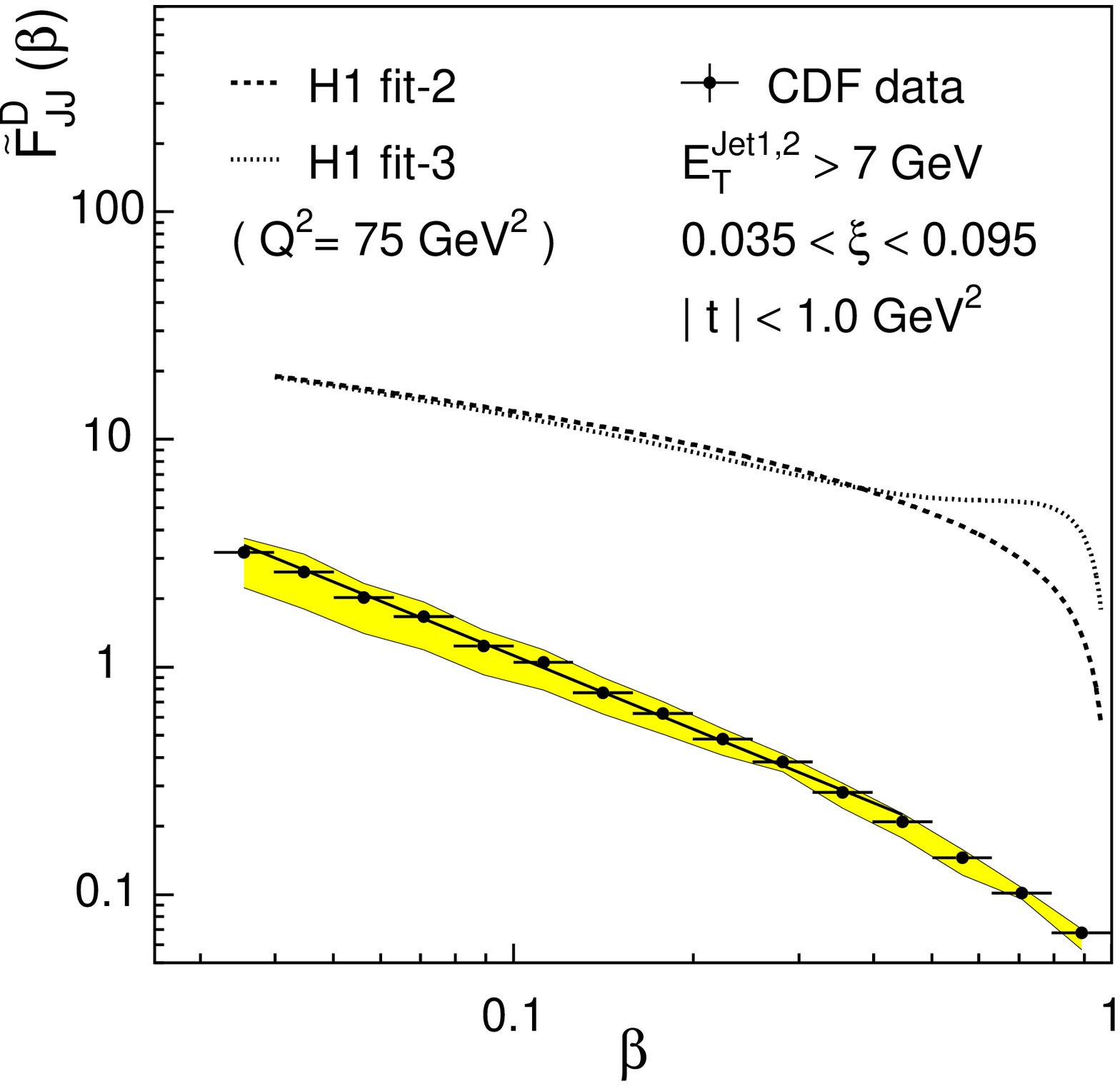}}
\renewcommand{\baselinestretch}{0.95}
\caption{\label{fig:sd_jj}
Left: Diagram of diffractive dijet production. Right: Data $\beta$ distribution
(points) compared with expectations from DDIS by H1 (dashed
and dotted lines). The straight line is a fit to the data of
the form $\beta^{-n}$. The filled band represents the range
of data expected when different numbers of jets are used in 
evaluating $\beta$. The normalization uncertainty of the
data is $\pm25$\%.}
\end{center}
\end{figure}

Diffractive events in $\bar{p}p$ collisions are characterized by 
the presence of a leading proton or antiproton which remains intact,
and/or a rapidity gap, defined as a pseudorapidity\footnote{The 
pseudorapidity $\eta$ of a particle is defined 
as $\eta \equiv -\ln (\tan \theta/2)$,
where $\theta$ is the polar angle of the particle with respect to 
the proton beam direction.} region devoid of particles. 
Diffractive events involving hard processes (``hard diffraction''), 
such as production of high $E_T$ jets (see 
Fig.~\ref{fig:diff_diagram}), have been studied extensively to
understand the nature of the exchanged object, the Pomeron,
which in QCD is a color singlet entity with vacuum quantum numbers. 
One of the most interesting questions in hard diffractive processes
is whether or not they obey QCD factorization, in other words, 
whether the Pomeron has a universal, process independent, parton 
distribution function (PDF). Results on diffractive deep 
inelastic scattering (DDIS) from the $ep$ collider HERA show that 
QCD factorization holds in DDIS. However, single diffractive (SD) rates 
of $W$-boson \cite{CDF_W}, dijet \cite{CDF_jj}, $b$-quark \cite{CDF_b} 
and $J/\psi$ \cite{CDF_jpsi} production relative to non-diffractive 
ones measured at CDF are $\cal{O}$(10) lower 
than expectations from PDFs determined at HERA, 
indicating a severe breakdown of factorization in hard diffraction 
between Tevatron and HERA.
The suppression factor at the Tevatron  relative to HERA 
is approximately equal in magnitude to that measured in
soft diffraction cross sections relative to Regge theory predictions based on
Regge factorization. The suppression relative to  predictions based on DDIS 
PDFs is illustrated in Fig.~\ref{fig:sd_jj}, which  shows the 
``diffractive structure function'' $F_{jj}^D$ measured at 
CDF by using diffractive dijet data with a leading antiproton detected in 
Roman Pots \cite{CDF_SDjj_1800, CDF_SDjj_630}. The $\tilde{F}_{jj}^D$ 
(integrated over antiproton momentum loss $\xi$ and four momentum 
transfer squared $t$) was obtained 
as a function of $\beta$, the momentum fraction of the parton in the Pomeron, 
$\beta=x_{\bar{p}}/\xi$ ($x_{\bar{p}}$ is $x$-Bjorken 
of the parton in the antiproton, see Fig.~\ref{fig:sd_jj}),
by measuring the ratio of diffractive to non-diffractive dijet
rates and using the known leading order PDFs of the proton. The 
measured suppression of $F_{jj}^D$ relative to the expectation from the 
H1 PDFs is 
approximately equal to that observed in soft diffraction.
CDF has also studied dijet events with a double Pomeron exchange 
(DPE) topology (Fig.~\ref{fig:diff_diagram}) using the Roman Pot 
trigger sample at $\sqrt{s}=1800$ GeV \cite{CDF_DPEjj}. By 
measuring the ratio of DPE to SD dijet rates ($R^{DPE}_{SD}$)
and comparing it with that of SD to ND rates ($R^{SD}_{ND}$), a breakdown
of QCD factorization was observed as a discrepancy of the double ratio
$D = R^{SD}_{ND}/R^{DPE}_{SD} = 0.19\pm0.07$ from unity.

\section{Run II Diffraction Measurements} 
In Run II, being currently under way, CDF plans to study various
topics on diffraction, including $Q^2$ and $\xi$ dependence of
$F_{jj}^D$ in SD, gap width dependence of $F_{jj}^D$ in DPE, 
production of exclusive dijet, heavy flavor and low mass states 
in DPE, and dijets with a large gap in-between jets. 
Two recently installed  ``Miniplug'' (MP) 
calorimeters cover the region $3.5<|\eta|<5.1$, 
and 7 stations of scintillation counters, called Beam Shower Counters (BSC), 
mounted around the beam pipe, extend the coverage to the 
very forward region of $5.5<|\eta|<7.5$. The Roman Pots (RP)
used in Run I were re-installed and are being operated to trigger
on leading antiprotons in the kinematic range $0.02<\xi<0.1$
and $0<|t|<2$ GeV$^2$. 

\begin{figure}[ht]
\begin{center}
\epsfxsize=6cm\epsfbox{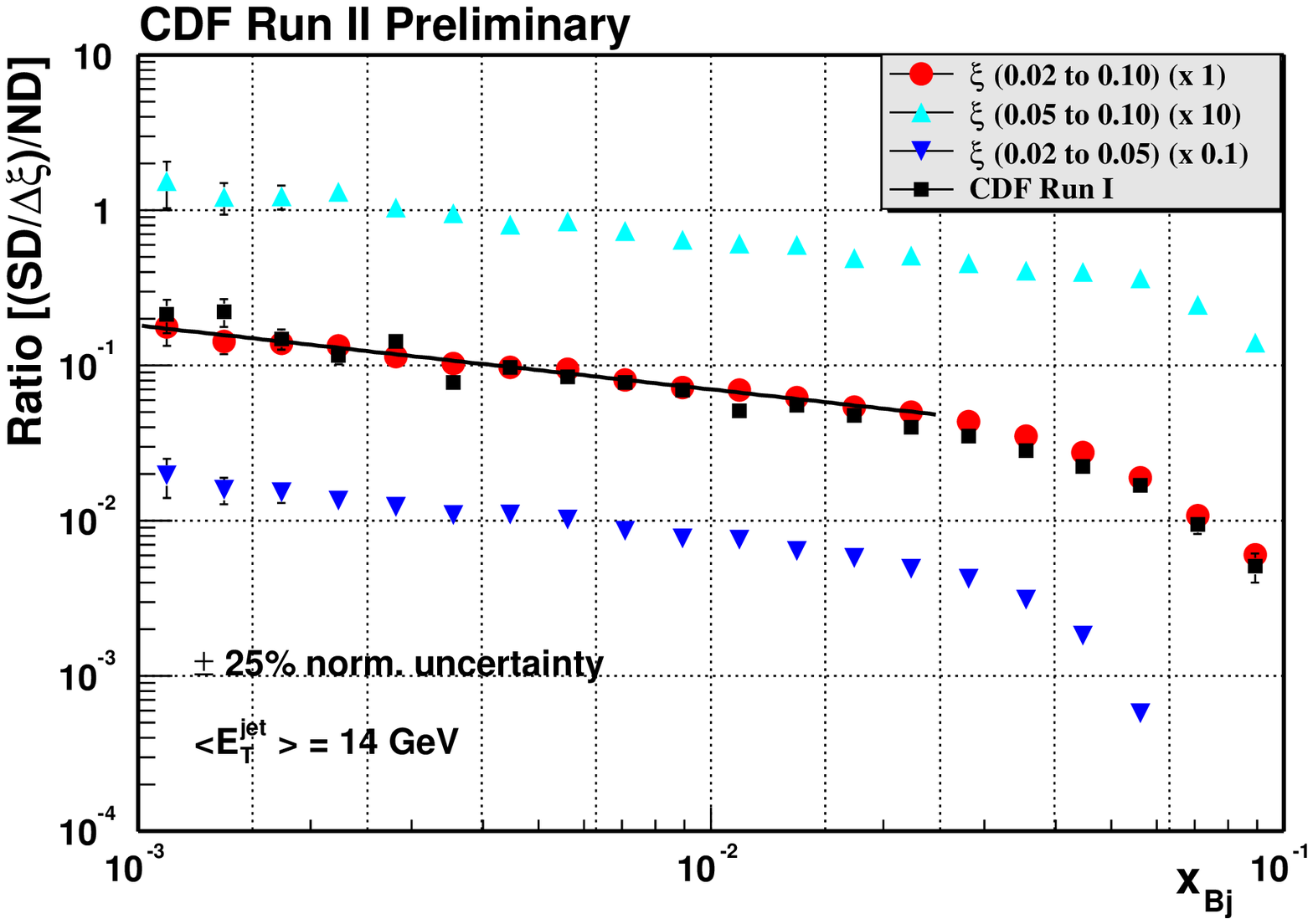}
\hspace{-0.3cm}\epsfxsize=6.0cm\epsfbox{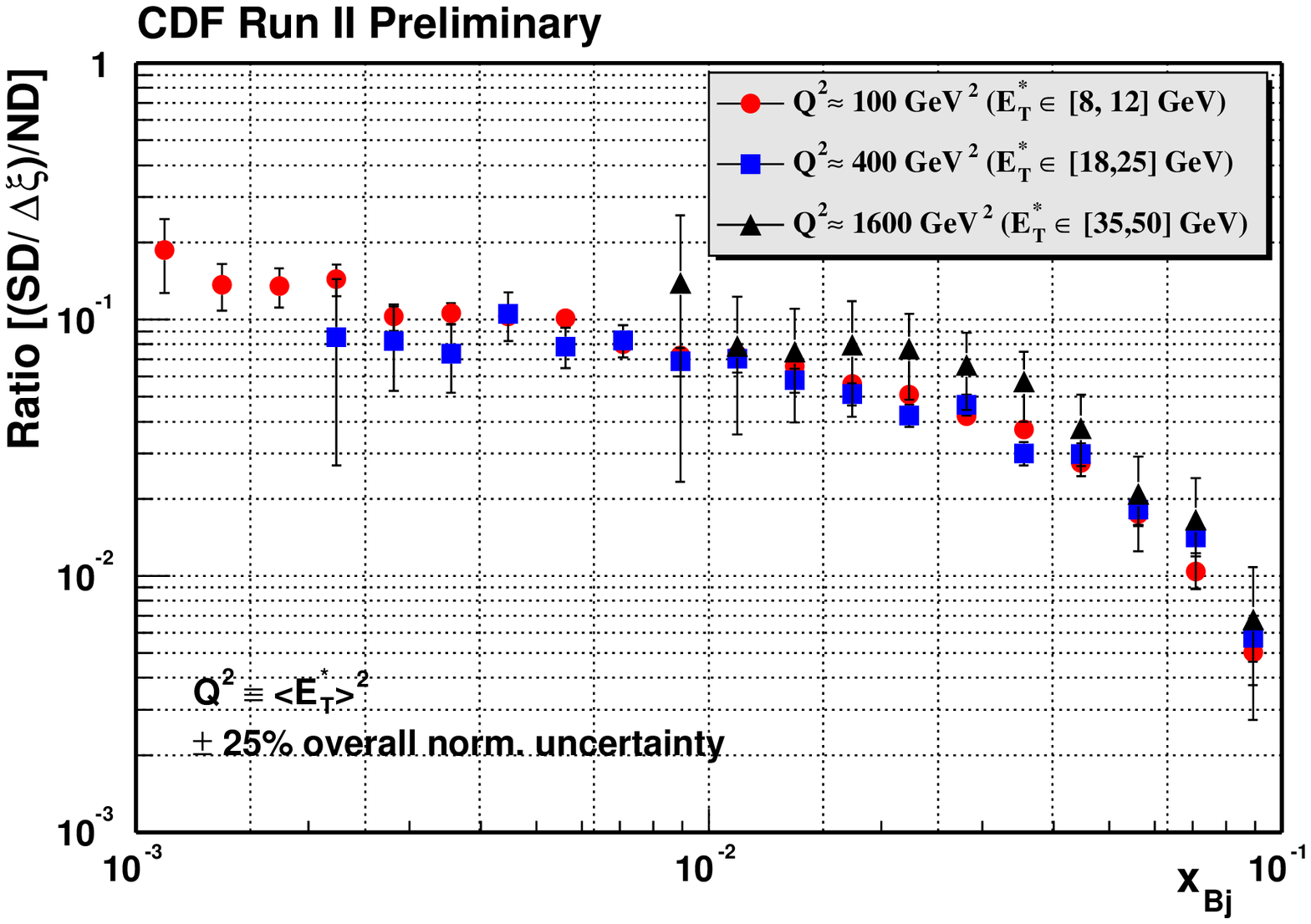}
\renewcommand{\baselinestretch}{0.95}
\caption{\label{fig:r_xbj}
Ratio of diffractive (per unit $\xi$) to non-diffractive dijet 
rates as a function of $x$-Bjorken of the parton in the antiproton. 
Left: ratios for different $\xi$ intervals 
(circle:$0.02<\xi<0.1$, downward triangle:$0.02<\xi<0.05$, 
upward triangle:$0.05<\xi<0.1$) for average dijet $E_T$ of 14 GeV,
compared with Run I measurement (square). Right: ratios for different 
$E_T^*$ (average jet $E_T$) intervals
(circle:$8<E_T^*<12$ GeV, square:$18<E_T^*<25$ GeV, 
triangle:$35<E_T^*<50$ GeV). $Q^2$
in this plot is defined as $Q^2=\langle E_T^* \rangle^2$.}
\end{center}
\end{figure}

\section{Diffractive Dijet Production in Run II} 
Triggering on a leading antiproton in the RP in conjunction with 
at least one calorimeter tower with $E_T>5$ GeV, a study of 
diffractive dijet events has been performed. From a sample 
of 352K triggered events, about 15K SD dijet events with dijets 
of corrected $E_T>5$ GeV in the range $0.02<\xi<0.1$ were 
obtained. The $\xi$ (fractional momentum loss of antiproton)
was measured by using all calorimeter information. Using a 
non-diffractive dijet sample triggered on the same calorimeter
tower requirement, the ratio of diffractive to non-diffractive
dijet rates was measured as a function of $x$-Bjorken of the parton 
in the antiproton, as shown in Fig.~\ref{fig:r_xbj}.
This figure shows that (i) the ratio observed with Run II data
in approximately the same kinematic region as in Run I reproduces the Run I 
results, and (ii) there is no appreciable $\xi$ dependence in the ratio,
as already seen in Run I. Measurement of the $\xi$ dependence at still
lower $\xi$ values ($\xi<0.02$) is one of our Run II goals 
and is being currently under study. Preliminary results of the 
$Q^2$ dependence of the ratio, where $Q^2$ is defined 
as the square of average value of the mean dijet $E_T$, are 
shown in Fig.~\ref{fig:r_xbj}. No
significant $Q^2$ dependence is observed, indicating that 
the Pomeron evolves with $Q^2$ in a similar way as the
proton.

\section{Dijet Production by Double Pomeron Exchange in Run II} 
\begin{figure}
\begin{center}
\epsfxsize=4.0cm\epsfbox{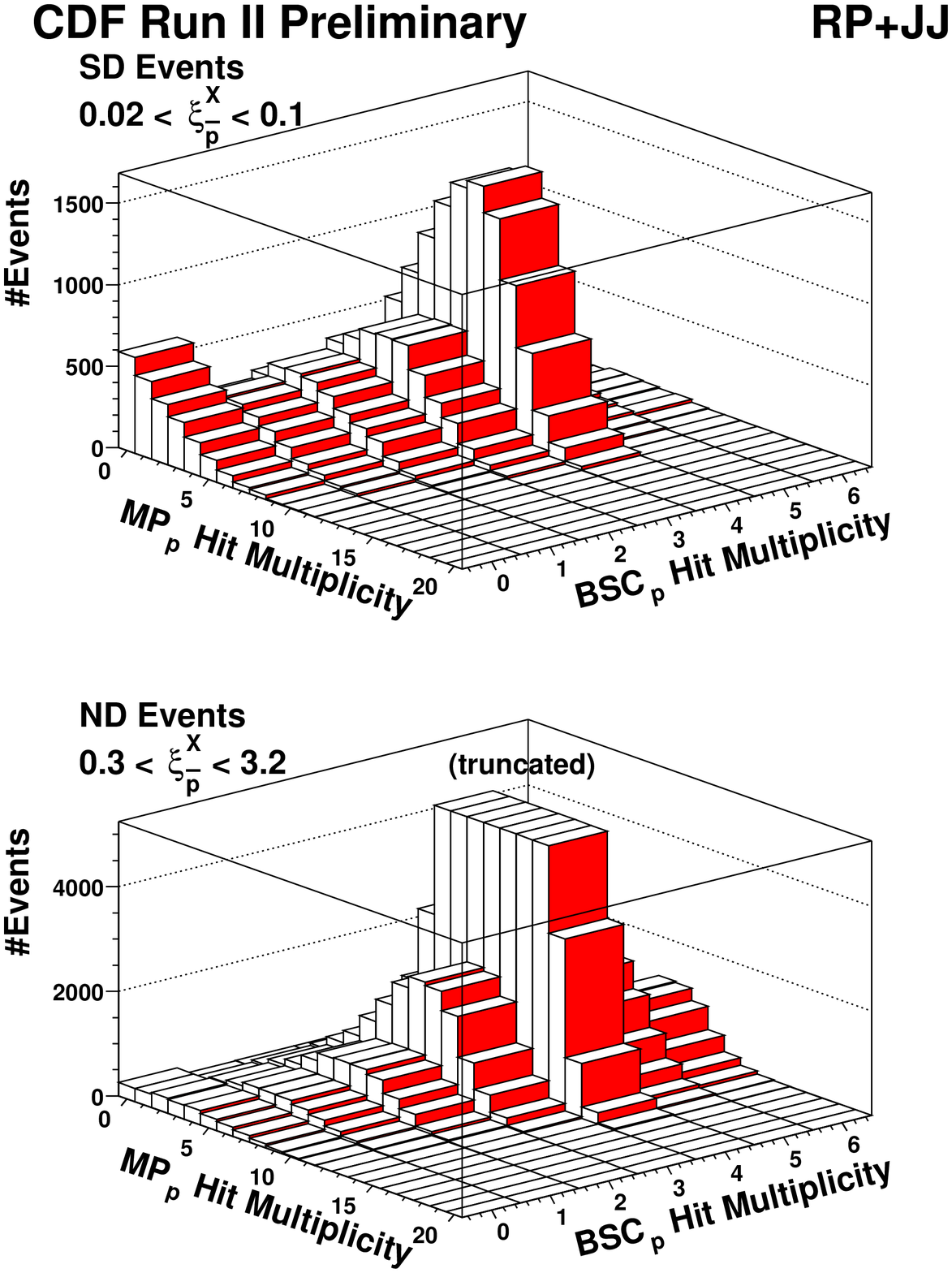}
\epsfxsize=3.7cm\epsfbox{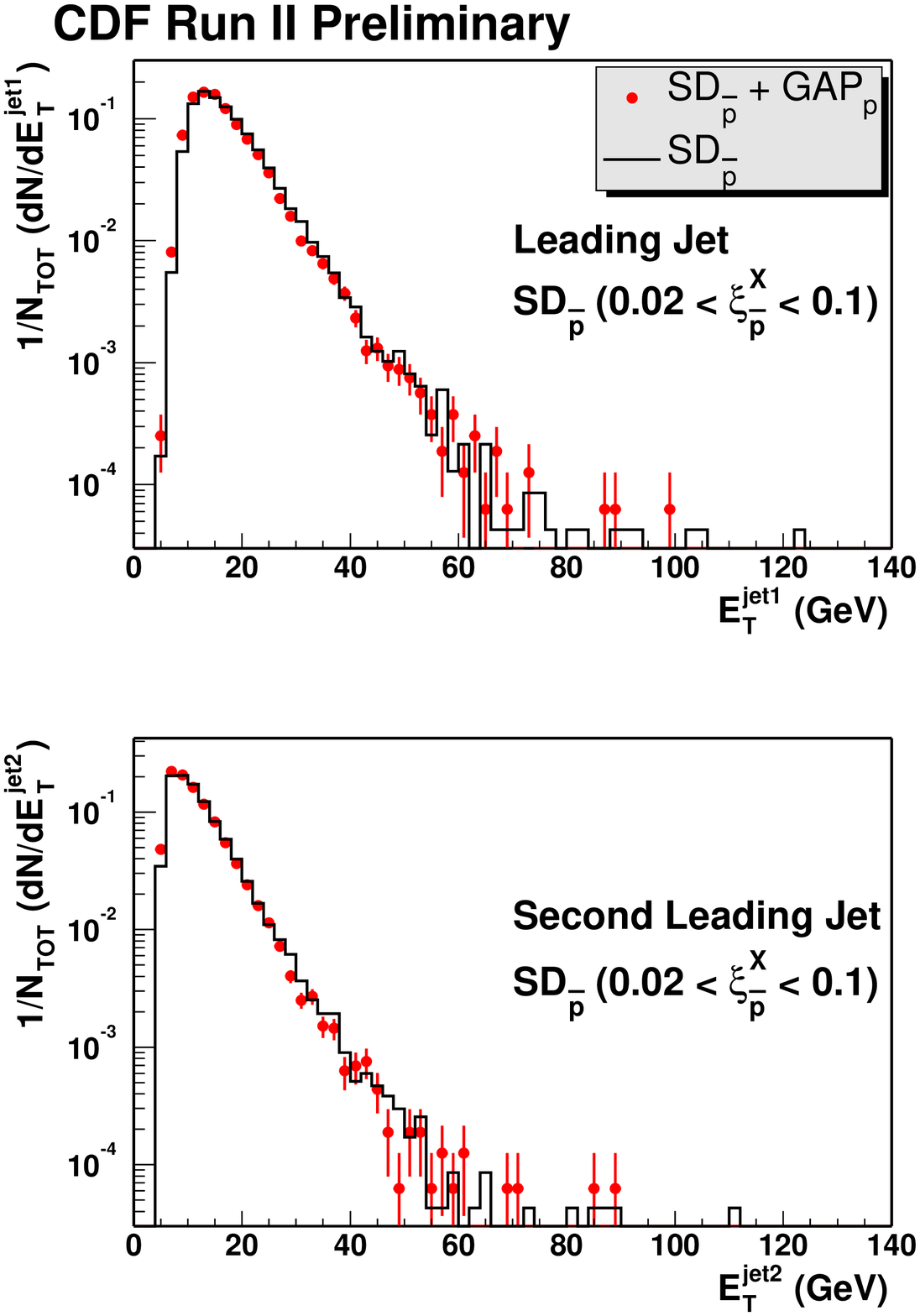}
\raisebox{2.8cm}{\epsfxsize=3.6cm\epsfbox{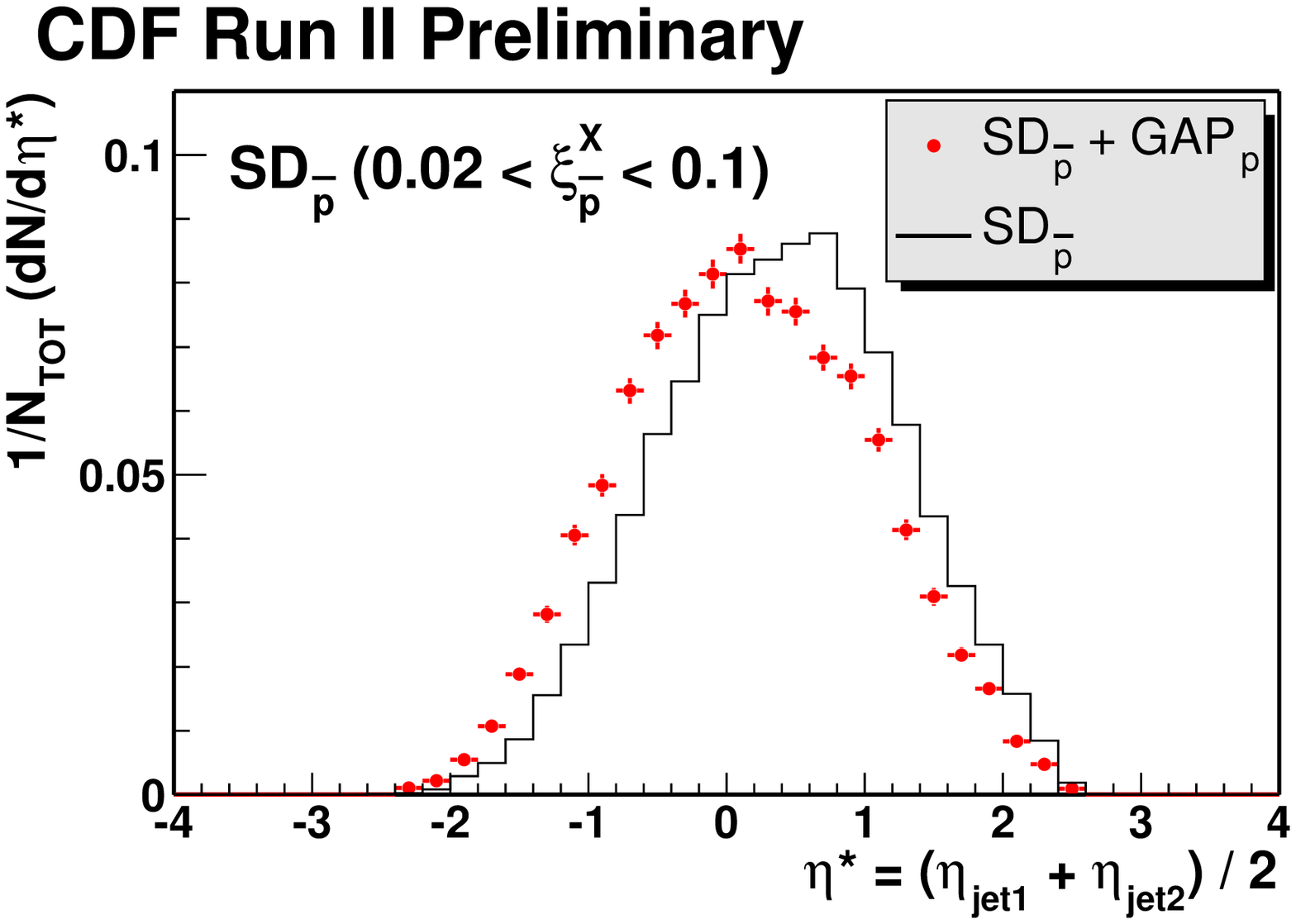}}\vspace{-2.6cm}\\
\hspace{8.1cm}\epsfxsize=3.6cm\epsfbox{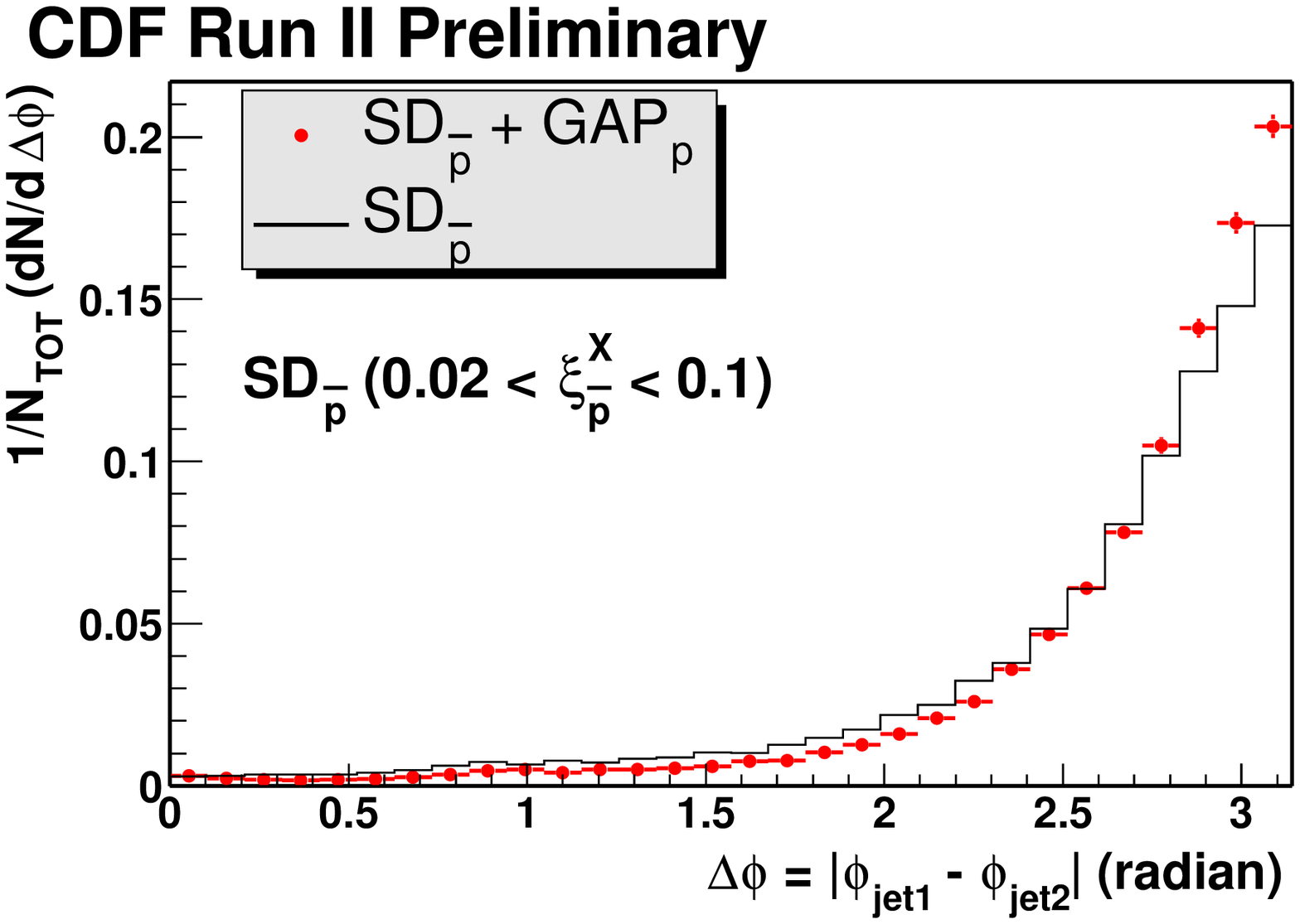}
\renewcommand{\baselinestretch}{0.95}
\caption{\label{fig:dpe}
Left: MP hit multiplicity versus BSC hit counter multiplicity on the
proton side in SD trigger sample. Events in the region 
$0.02<\xi<0.1$ ($0.3<\xi<3.2$) are shown in the top (bottom)
plot. The $\xi$ is calculated from calorimeter tower information. 
Middle and Right: $E_T$ distributions of the leading (middle-top) 
and second to leading (middle-bottom) jets, mean $\eta$ (right-top) 
and $\Delta\phi$ (right-bottom) of the two leading jets in DPE dijet 
events (points) compared with those of the SD dijet sample (histograms). 
All distributions are normalized per unit area.}
\end{center}
\end{figure}

For a study of DPE dijets in Run II, a dedicated trigger has been implemented
that requires a rapidity gap in the BSC in the outgoing proton 
direction in addition to the presence of a leading antiproton  in the RP
and a single calorimeter tower of $E_T>5$ GeV. 
The requirement of a BSC gap on 
the proton side enhances the DPE signal, as can 
be seen in the two-dimensional LEGO plot of MP versus BSC 
hit multiplicity of SD dijet events (Fig.~\ref{fig:dpe}). 
Offline, requiring in 
addition a gap in the proton-side MP, 
we obtained about 16K dijet events (about 100 times more data 
than in Run I), which are qualitatively consistent with DPE dijets. 
Figure~\ref{fig:dpe} (middle and right) shows  the $E_T$, mean $\eta$ 
and azimuthal angle difference $\Delta\phi$
of the two leading jets for the DPE candidate events (points). As seen 
in Run I DPE data, the $E_T$ distributions look similar to those of SD 
dijets (histograms), while the mean $\eta$ and $\Delta\phi$ 
show that the DPE dijets are more central and more back-to-back. 

\end{document}